\documentclass{aastex}
\usepackage{emulateapj5}
\usepackage{onecolfloat5}
\usepackage{natbib}
\bibpunct{(}{)}{;}{a}{,}{,}

\input epsf
\def\plotone#1{\centering \leavevmode
\epsfxsize= 1.0\columnwidth \epsfbox{#1}}
\def\plottwo#1{\centering \leavevmode
\epsfxsize= 0.75\columnwidth \epsfbox{#1}}



\long\def\comment#1{}

\def\W2{{\cal W}}

\def\be{\begin{equation}}
\def\ee{\end{equation}}
\def\bea{\begin{eqnarray}}
\def\eea{\end{eqnarray}}
\def\sm{{\rm M}_\odot}

\def\Mpc{\,{\rm Mpc}}
\def\mpc{\,{\rm Mpc}}

\def\cmm2{{\,\rm cm^{-2}}}
\def\cm2{{\,{\rm cm}^2}}
\def\cmm3{{\,{\rm cm}^{-3}}}
\def\gcmm3{{\,{\rm g\,cm^{-3}}}}
\def\kms{\,{\rm km\,s^{-1}}}

\def\fun#1#2{\lower3.6pt\vbox{\baselineskip0pt\lineskip.9pt
  \ialign{$\mathsurround=0pt#1\hfil##\hfil$\crcr#2\crcr\sim\crcr}}}

\def\muk{\mu{\rm K}}

\lefthead{Dor\'e et al.} \righthead{Gravitational potential from
peculiar velocities and weak lensing measurements}

\begin{document}
\bibliographystyle{apj}
\twocolumn[
\submitted{To be submitted to ApJ Letters}
\title{Gravitational Potential Reconstruction
from Peculiar Velocity and Weak Lensing Measurements}
\author{Olivier Dor\'e$^{1}$, Lloyd Knox$^{2}$ and Alan Peel$^2$}
\affil{$^{1}$ Institut d'Astrophysique de Paris, 98bis boulevard Arago, 75014
Paris, FRANCE, email: dore@iap.fr}
\affil{$^{2}$ Department of Physics, University of California, Davis, CA 95616, USA,
email: lknox@ucdavis.edu}

\begin{abstract}
We present an analytic method for rapidly forecasting the accuracy
of gravitational potential reconstruction possible from
measurement of radial peculiar velocities of every galaxy cluster
with $M> M_{\rm th}$ in solid angle $\theta^2$ and over redshift
range $z_{\rm min}<z <z_{\rm max}$.  These radial velocities can
be determined from measurement of the kinetic and thermal
Sunyaev--Zeldovich effects. For a shallow survey with $0.2 < z <
0.4$, coincident with the SDSS photometric survey, one mode of the
gravitational potential (on length scales $\ga 60\mpc$) can be
reconstructed for every $\sim 8$ cluster velocity determinations.
Deeper surveys require measurement of more clusters per $S/N > 1$
mode. Accuracy is limited by the ``undersampling noise'' due to
our non--observation of the large fraction of mass that is not in
galaxy clusters. Determining the gravitational potential will
allow for detailed study of the relationship between galaxies and
their surrounding large--scale density fields over a wide range of
redshifts, and test the gravitational instability paradigm on very
large scales. Observation of weak lensing by large--scale
structure provides complementary information since lensing is
sensitive to the tangential modes that do not affect the velocity.
\end{abstract}

 ]

\section{Introduction}

Although there are a variety of techniques for measuring the
statistical properties of cosmological density fluctuations, we
know of only three types of observations from which the density
field itself may be reconstructed: weak lensing
\citep{mellier99,bartelmann01}, peculiar velocities
\citep{strauss95} and galaxy rotation vectors \citep{lee00}. A map
of the density field, combined with galaxy surveys, would be a
highly valuable aid to understanding the formation of galaxies and
clusters of galaxies. Numerical simulations could be performed
with realizations of initial conditions constrained by our
knowledge of the density field, allowing object--by--object
comparison between theory and observation, rather than solely
statistical comparison.  A map could also provide a guide to
observers who may, for example, wish to search for the luminous
tracers of the filamentary density structures.

Reconstruction of the density field from galaxy peculiar
velocities was pioneered by \citet{dekel90}. The radial component
of peculiar velocities of galaxies can be determined by
inferring the distance and then subtracting off the Hubble
flow contribution to the redshift.
Galaxy peculiar velocity determinations, because they rely on
distance determinations, are only useful at $z \la 0.1$.

Peculiar velocities of galaxy {\em clusters} determined from
observations of the Sunyaev--Zeldovich (SZ) effects
\citep{sunyaev80} do not rely on a distance determination. The
peculiar velocity signal arises from the cluster's radial motion
with respect to the cosmic microwave background (CMB).
High--resolution ($\la 1'$), multi--frequency observations of
galaxy clusters can be used to determine the radial velocities of
galaxy clusters with an accuracy that is independent of the
distance to the galaxy cluster\footnote{This independence only
fails for clusters at $z \la 0.2$ which suffer significant
confusion with primary CMB anisotropy due to their large angular
sizes.}. Errors as low as $100\kms$ may be achievable
\citep{holder02}. In addition to potential reconstruction,
velocity measurements will be useful for cosmological parameter
estimation \citep{peel02}.

Here we present a method for rapidly forecasting the accuracy of
density field reconstruction from peculiar velocity measurements
and apply it to surveys with various redshift ranges. We also show
how weak lensing observations can provide complementary
information \citep{mellier99,bartelmann01}.

Our analytic method for forecasting results assumes a uniform and
continuous velocity field map.  Even if this map were derived from
noiseless peculiar velocity measurements, there would still be an
effective noise contribution from the fact that most of the mass in the
Universe is not in galaxy clusters.  Fortunately, as we quantify below,
the contribution to the peculiar velocity variance from each logarithmic
interval in wavenumber $k$ drops as $k^{-1}$ for relevant scales, so this
noise from under--sampling is not overwhelmingly large.

We consider three different survey types labeled ``SDSS'',
``DEEP/VIRMOS''  and ``SZ'' with redshift ranges $0.2<z<0.4$,
$0.7<z<1.4$ and $0.2<z<2$ respectively.  Galaxy cluster redshifts
are required to convert SZ measurements into gravitational
potential maps. Thus, the redshift ranges of two survey types are
subsets of those of optical surveys SDSS (Sloan Digital Sky
Survey), DEEP II (Deep Extragalactic Evolutionary Probe;
\citet{davis02})  and VIRMOS (Visible-Infrared Multi-Object
Spectrographs; \citet{lefevre02}). The peculiar velocity surveys
we imagine could be realized by targeted multi--frequency SZ
follow--up on galaxy clusters identified in those optical surveys.
The ambitious, deep ``SZ" survey is modelled after proposed SZ
surveys which can locate galaxy clusters over a large $z$ range.
Again, the velocities can be determined from multi--frequency
follow up (if necessary).  The redshift determinations for this
last survey type are left as a currently unsolved challenge.

Radial peculiar velocities are directly sensitive to radial
gradients in the gravitational potential.  Gravitational lensing
is caused by tangential gradients in the gravitational potential;
thus, in principle, lensing provides complementary information for
potential reconstruction. However, we show below that lensing is
actually unlikely to add much to the density field reconstruction.
The reason is simple: lensing observations are two--dimensional,
whereas velocity observations are three--dimensional.  The
conclusion may be different if one considers lensing of many
source populations with differing redshift distributions
\citep{hu02b}.

\section{Forecasting Errors}

To begin, we assume that we have a continuous and uniform map of
the velocity field in a box of volume $V$.  The map is not
noise--free, but has white noise with finite weight--per--comoving
volume $w$; i.e, the variance of the error on the average velocity
in some sub-volume $V_1$ is $\sigma^2_{V_1}= 1/(wV_1)$.  For
specificity we set our fiducial model to be
$h=1-\Omega_m=\Omega_\Lambda=0.7$ and $\sigma_8 = 1$.

For the moment we ignore the fact we can only make measurements
along our past light cone and assume the map is of the velocity
field {\em today}.  We remove this oversimplification below, but
for now this idealization maintains homogeneity and allows a
completely analytic analysis of the potential reconstruction
errors in Fourier space. With the assumption of a potential flow
\citep{peebles93}\be {\vec v}_{\vec k} = i{\vec k\over H_0}
\Phi_{\vec k}(t_0)x(t) \ee where \be x(t)\equiv {2\over 3}{dD\over
da}\left(t\right){E(t)a^2(t)\over \Omega_m} \ee and \be E^2(t)
\equiv H^2(t)/H_0^2=\Omega_m/a^3(t)+\Omega_\Lambda. \ee The noise
and signal contributions to the variance of $\Phi_{\vec k}$ are
diagonal with diagonal entries given by \bea
N(k)& = &\left({H_0 \over k x_0}\right)^2/w \ \ {\rm and}\\
S(k) & = & P_\Phi(k)={9\over 4}\left({H_0 \over k}\right)^4\Omega_m^2
{2\pi^2\over k^3} \Delta^2(k)
\eea
respectively where $\Delta^2(k) \equiv k^3 P_\delta(k)/(2\pi^2)$ and
$P_\delta(k)$ is the matter power spectrum.

The square of the signal--to--noise ratio for the gravitational
potential map is thus
\bea
\label{eqn:sovern}
S(k)/N(k) &=& 1\left({w\over w_f}\right)\left({\Delta^2(k) \over
0.8}\right)\left({x_0 \over 1.1}\right)^2 \nonumber \\
& & \times\left({\Omega_m h\over0.21}\right)^2 \left({k\over 0.1 \ {\rm
Mpc}^{-1} }\right)^{-5} \nonumber \\
w_f^{-1} &\equiv& \left(100\kms\right)^2\left(64 \Mpc \right)^3
\eea where $x_0 \equiv x(t_0) = 1.1$ and
$\Delta^2(0.1\Mpc^{-1})=0.8$ for our fiducial model. We therefore
expect measurements with $S/N > 1$ for every mode with $2\pi/L \ga
k \ga 0.1 \Mpc^{-1}$ where $L^3=V$. Further, there are many of
these modes for a box of size L; the number in Fourier--space
volume $(\Delta k)^3$ is $4\times 10^3\left((k/0.1\Mpc^{-1})
(L/1000\Mpc)(\Delta k/k)\right)^3$.

\begin{figure}
\plottwo{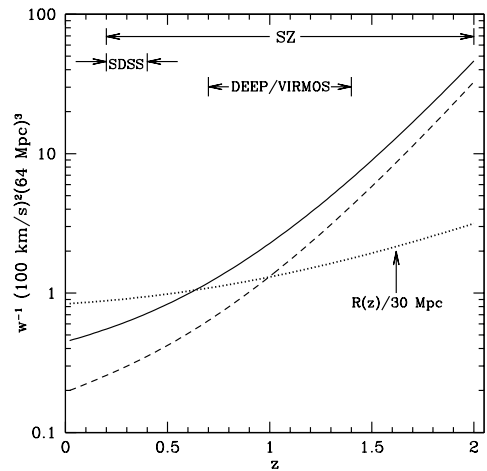}
\caption[]{\label{fig:velweight} Inverse weight--per--solid angle,
$w^{-1}$, from under--sampling noise (dashed line) and from
under--sampling noise plus a $100 \kms$ error on each galaxy
cluster added in quadrature (solid line). The dotted line is
$R(z)/30\Mpc$ where $4/3 \pi R^3(z) \bar n(z)= 1$ defines $R(z)$
and $\bar n(z)$ is the comoving average number density of
clusters.}
\end{figure}

We now address our value of $w_f$ above.  A fundamental limit on
$w$ is set by the small fraction of mass in galaxy clusters. The
comoving number density of galaxy clusters with lower mass
threshold $10^{14}h^{-1}\sm$ at $z=1$ is about one per
$(64\Mpc)^3$ volume \citep{holder01} which roughly corresponds to
the volume of a $40\Mpc$ radius sphere.  However, a sphere
containing this limiting mass in a uniform background only has a
comoving radius of about $R=9$ Mpc. Estimating the velocity field
averaged over $R=40\Mpc$ from just one sample of the velocity
field measured over $9\Mpc$ introduces an under--sampling error
with variance (at $z=1$) of $\langle (v_{40}-v_9)^2 \rangle =
(117\kms)^2$.  Thus we cannot do better than
$w^{-1}=(117\kms)^2(64\Mpc)^3$ at $z=1$. The dashed line in
Fig.~\ref{fig:velweight} shows the redshift dependence of this
under--sampling noise which is due to evolution in $\bar n$.
According to \citet{holder02} and \citet{Nagai02} there is also a
fundamental limit on how well one can infer the peculiar velocity
of a galaxy cluster from the (highly non--uniform) velocities of
the gas, of about $100\kms$. The result of adding these errors in
quadrature is the solid line of Fig.~\ref{fig:velweight}. Wiener
filtering can reduce the error, though we do not employ it here.

The angular resolution, sensitivity and frequency coverage
required to achieve a velocity measurement with a given precision
have not yet been worked out in detail. From prior work
\citep{haehnelt96,holzapfel97,kashlinsky00,aghanim01}, which
ignores the velocity substructure of the gas, we find measurements
with errors of $8 \mu K$ translate into $\sigma_v \simeq 100
\kms$, well--matched to the Holder/Nagai limit. A multi--mode
detector on a large telescope with sensitivity of $100 \muk
s^{1/2}$ could achieve $8 \muk$ on 20 clusters in 1 hour, or
16,000 in a month.

We now consider observations of the radial component of peculiar
velocities on our past light cone. The isotropic but inhomogeneous
geometry suggests the mode decomposition \be \Phi(\hat\gamma,r) =
\sum_{lm}\int_0^\infty dk k^2 \tilde \Phi_{lm}(k) j_l(kr)
Y_{lm}(\hat \gamma) \ee as used by e.g. \citet{stebbins96}.  The
line--of--sight radial velocity generated by this potential is \be
v_r(\hat \gamma,r) = ix(r)/H_0 \sum_{lm}\int_0^\infty dk k^3
\tilde \Phi_{lm}(k) j_l'(kr) Y_{lm}(\hat \gamma)
\label{eqn:Phi2vr} \ee where we have changed our independent
variable from $t$ to conformal distance, $r$ and $j_l'(y) \equiv
d/dy\ j_l(y)$. Note that all the time--dependence on the
right--hand side is carried by $x(r)$, thus we are discussing
reconstruction of the gravitational potential on our past light
cone {\em extrapolated by linear theory to what it would be
today}.

Our discretized model of the data is \be v_i = \sum_{blm}
A_v(i;b,l,m)\tilde \Phi_{blm} + n_i \ee where $n$ is the error
(whose statistical properties were just discussed above), $A_v$ is
implicitly defined by Eq.~\ref{eqn:Phi2vr} and $b$ enumerates the
$k$ bins: \be \tilde \Phi_{blm} \equiv {1\over \Delta k_b}
\int_{k_b-\Delta k_b/2}^{k_b+\Delta k_b/2}dk \tilde \Phi_{lm}(k).
\ee  The minimum--variance estimate of $\tilde \Phi_{blm}$ is \bea
\hat {\tilde \Phi}_{blm}& = & \sum_{b'l'm'ii'} W_{v
\Phi}^{-1}(blm,b'l'm') \nonumber \\
&\times& A_v(i;b',l',m')N^{-1}_{ii'} v_{i'} \eea where $N_{ij} =
\langle n_i n_j \rangle$, \be W_{v\Phi}^{-1}(blm,b'l'm') = \langle
\delta \Phi_{blm} \delta \Phi_{b'l'm'}\rangle \ee is the noise
covariance matrix of $\hat{\tilde \Phi}_{blm}$ given by \be
W_{v\Phi}(blm,b'l'm') = \sum_{ii'} A_v(i;b,l,m) N^{-1}_{ii'}
A_v^*(i';b',l',m') \ee and $\delta \tilde \Phi_{blm} \equiv \hat
{\tilde \Phi}_{blm} - \tilde \Phi_{blm}$.

With the uniform sampling assumption, $W_{v \tilde \Phi}$ for
reconstructing $\Phi_{blm}$ is given by
\bea
& W_{v \tilde \Phi}(blm,b'l'm') = {\Delta k_b  \over H_0} {\Delta k_{b'}
\over H_0} \delta_{ll'}\delta_{mm'} k_b^3 k_{b'}^3 \nonumber &\\
& \times\int dr r^2 x^2(r) w(r) j_l'(k_br)j_l'(k_{b'}r). & \eea
The matrix is block--diagonal with zero entries for any elements
with $l \ne l'$ or $m \ne m'$.  As one expects from statistical
isotropy $W_{v\tilde \Phi}$ does not depend on $m$.

The signal matrix is diagonal, though the effect of binning the
uncorrelated $k$ modes is to reduce the variance by $\Delta k$:
$\langle \tilde \Phi_{blm} \tilde \Phi_{blm} \rangle = P_{\tilde
\Phi}(k)/\Delta k$.  In the results section we compare signal and
noise by plotting $k^5 \Delta k/ W_{v \tilde \Phi}$ where the
$k^5$ factor makes a dimensionless quantity and the $\Delta k$
makes the corresponding signal quantity, $k^5 P_{\tilde \Phi}(k) =
(4\pi)^2 $ $\times k^3 P_\Phi(k)$, independent of $\Delta k$.

The weight matrix for potential reconstruction from a weak lensing
convergence map with uniform weight per solid angle $w$: \bea
\label{eqn:lensW} & & \lefteqn{ W_{\kappa \tilde \Phi}(blm,b'l'm')
=
\delta_{ll'}\delta_{mm'}wl^2(l+1)^2 }  \nonumber\\
&  & \times \Delta k_b k_b^2 \Delta k_{b'}
k_{b'}^2I_l(k_br_s,\theta_s)I_l(k_{b'}r_s,\theta_s) \eea where \be
I_l(kr_s,\theta_s) = \int_0^{r_s}{dr \over rr_s}
\left(r_s-r\right)j_l(kr){D(\eta_0-r) \over a(\eta_0-r)}
W(kr\theta_s)\ee \citep{stebbins96}, and
$W(kr\theta_s)=2J_1(kr\theta_s)/(kr\theta_s)$ \citep{jain97}
incorporates the fact that we smooth the signal over radius
$\theta_s = 2.5'$.  Again, this is block--diagonal in $l$ and $m$.
However, the structure in the radial wave--number $k$ is highly
degenerate. For each $l, m$ sub--matrix, the $k,k'$ dependence can
be written as an outer product of a single vector.  Such matrices
have only one non--zero eigenvalue, a manifestation of the
inability of lensing measurements to distinguish between different
radial modes.

\section{Results}

\begin{figure}
\plotone{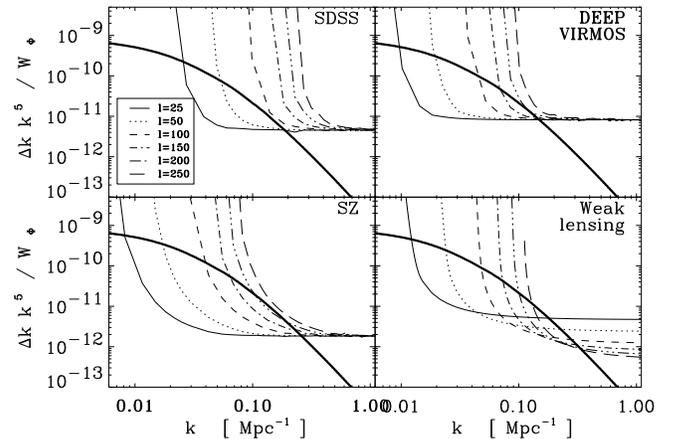}
\caption[]{\label{fig:diagonal} Dimensionless diagonal noise
variance ($\Delta k k^5 / W_{X\tilde\Phi}$) for binned modes of
the gravitational potential with bin widths $\Delta k = k$ for
several values of $l$, for our three velocity surveys ($X=v$) and
the lensing survey $(X=\kappa)$. The expected signal variance,
$k^5 P_\Phi(k)$ is also shown for comparison.}
\end{figure}

In Figure~\ref{fig:diagonal} we compare signal and noise for
$\Delta k = k$ for several values of $l$ assuming $w(z)$ as shown
in Fig.~\ref{fig:velweight}. One can see that $S/N > 1$
measurements are possible on large spatial and angular scales. The
sharp low $k$ cutoffs are an effect of the finite size of the
surveys:  $k_{\rm min}=l/L$ where $L$ is the radial depth of the
survey since $j_l(kr) \simeq 0$ for $k<l/r$. The velocity results
can be understood quantitatively by multiplying $S/N$ from
Eq.~\ref{eqn:sovern} by the number of modes in a bin of width
$\Delta k$: $\Delta k L/(2\pi)$.

Now we turn to the weak lensing panel of Fig.~\ref{fig:diagonal}.
We assume a noise level of $w=9 \times 10^9$ which can be achieved
with a galaxy number density of $n = 30\ \rm{gal.\ arcmin}^{-2}$
and intinsic galaxy ellipticity of $\sigma_{\epsilon}=0.2$
\citep{kaiser98,vanWaerbeke99}.  We further assume a negligible
width to the distribution of these galaxies centered at $z_s=1$.
The results can be understood analytically from
Eq.~\ref{eqn:lensW} and the approximation \citep{stebbins96} $I_l
= \sqrt{\pi/(2l)}(1/l-1/(kr_s))$ for $kr_s > l$ and $I_l = 0$
otherwise.  Unlike with velocity the higher $l$ values are
reconstructed with less noise due to the sensitivity of the
convergence to tangential gravitational potential variations.

\begin{figure}
\plotone{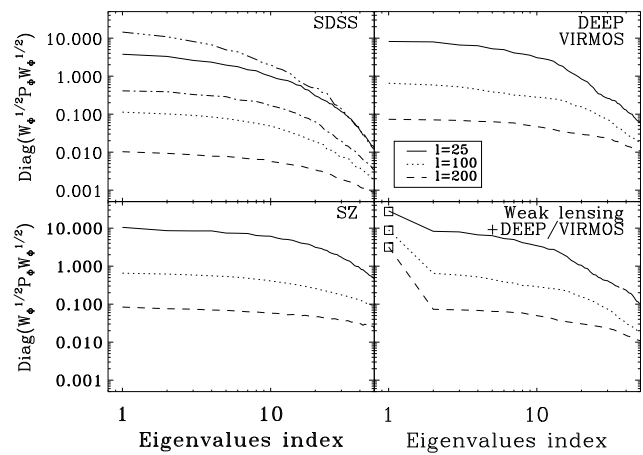}
\caption[]{\label{fig:eigenvalues} Signal--to--noise eigenvalue
spectra for individual angular modes. The boxes show the single
non--zero eigenvalues of the weak lensing survey. The SDSS panel
includes additional curves for $l=10$ (top curve) and $l=50$.
}
\end{figure}

Above we have plotted $k_b^6/W_{v \Phi}(blm,blm)$ as an indicator
of the variance expected on a measurement of $k_b^3 \Phi_{blm}$.
However, the variance of the error on this mode is actually given
by $k_b^6W^{-1}_{v\ \Phi}(blm,blm)$. This inverse is not
well--defined because of eigenmodes with zero eigevnalues.    In
Fig.~\ref{fig:eigenvalues} we plot the eigenvalues of the
signal--to--noise matrix $W_{v\tilde \Phi}^{1/2}P_{\tilde
\Phi}/\Delta k W_{v \tilde \Phi}^{1/2}$ (e.g., \cite{bond95}).

The total weight from a combined weak lensing and peculiar
velocity survey is simply given by adding the individual weights.
The eigenvalue spectrum for this combination is identical to the
peculiar velocity survey eigenvalue spectrum, except for the
increased value of one eigenvalue per $l,m$ mode.

\section{Discussion}

The largest $l$ value of modes with $S/N > 1$, $l_{\rm max}$,
entails a minimum survey size of $\theta_{\rm
min}=5^\circ(80/l_{\rm max})$. From Fig.~\ref{fig:eigenvalues} we
see $l_{\rm max} \simeq 80$ and therefore $\theta_{\rm min}\sim
5^\circ$ for the two deeper surveys and $l_{\rm max} \simeq 35$
and therefore $\theta_{\rm min}\simeq 10^\circ$ for SDSS.

There are many advantages to the shallower SDSS survey.  The
spatial number density of clusters is higher, reducing the
under--sampling noise.  Despite the increased value of
$\theta_{\rm min}$, the clusters in solid angle $\theta_{\rm
min}^2$ are actually fewer for the shallow survey (358) than for
the deeper surveys (500 and 1000). This is an advantage since at
fixed detector sensitivity the requisite observing time is
proportional to the number of clusters, {\em not} the survey solid
angle. Nearer clusters are easier to observe in the optical and in
X--rays, so redshift and temperature determinations will take less
time. Finally, a nearby map would be easier to compare to other
tracers of the density field than a more distant map would be.

Increasing the survey solid angle $\theta^2$ not only reduces
$l_{\rm min}$, but also increases spectral resolution (i.e.,
decreases $\delta l = 2\pi/\theta = 12(30^\circ/\theta)$) and
increases the number of angular modes with $S/N > 1$:
\be N_\theta = {\theta^2 \over 4\pi}\sum_{l=l_{\rm min}}^{l_{\rm
max}}\left(2l+1\right)=27\left(\theta \over
30^\circ\right)^2\left(l_{\rm max} \over 35\right)^2-\pi. \ee Thus
for a $\theta = 30^\circ$ survey with $0.2 < z < 0.4$ the total
number of modes with $S/N \ga 1$ is roughly $N_\theta \times
15\simeq 360$.  Covering the entire $\pi$ steradians of the SDSS
survey would require measuring 36,000 clusters and would deliver
$\sim 4,500$ modes.

\section{Conclusions}

We have shown that galaxy cluster peculiar velocities can be used
to make high signal--to--noise maps of the gravitational
potential, and therefore matter density, on very large scales. A
limiting source of uncertainty in these maps is what we have
called the under--sampling noise due to the fact that only a small
fraction of mass is in clusters.  The under--sampling noise can be
reduced by lowering the mass threshold but the lower the mass of
the cluster, the harder the measurement of its peculiar velocity.
Much more work is needed to understand the demand on experimental
resources and to optimize observing strategies. We have made
several arguments that favor shallower surveys over deeper ones.

The $\Phi$ and density maps will have numerous applications.
Cross--correlations with various tracers of the density field (red
galaxies, blue galaxies, infrared galaxies, lensing of galaxies,
lensing of the CMB, integrated Sachs--Wolfe effect on CMB, ...)
will be of great interest.  The greatest drawback of the maps is
their inability to probe scales below the typical separation
between clusters.

\acknowledgments

We thank G. Holder for his encouragement and for sharing the $\bar
n(z)$ calculation used in \citet{holder01} and L. van Waerbeke and
S. Colombi for several useful conversations. O.D. acknowledges
financial support from ``Soci\'et\'e de Secours des Amis des Sciences''.

\bibliography{cmb3}
\end{document}